\documentclass[12pt]{article}
\usepackage{amsfonts}
 \usepackage{fancyhdr}
 \usepackage{graphicx}
 \usepackage{fancybox}
 \usepackage{fancyvrb}
\begin{document}

\def\ba{\begin{eqnarray}}
\def\ea{\end{eqnarray}}

\begin{titlepage}
\title{{\bf Lagrange formulation of the symmetric teleparallel gravity}}
\author{ M. Adak  \\
 {\small Department of Physics, Faculty of Arts and Sciences,} \\
 {\small Pamukkale University, 20100 Denizli, Turkey} \\
 {\small \tt madak@pamukkale.edu.tr} \\ \\
  M. Kalay  \\
 {\small Department of Physics, Faculty of Arts and Sciences,} \\
 {\small Pamukkale University, 20100 Denizli, Turkey} \\
  {\small \tt mkalay@pamukkale.edu.tr} \\ \\
   \"{O}. Sert \\
 {\small Department of Physics, Faculty of Arts and Sciences,} \\
 {\small Pamukkale University, 20100 Denizli, Turkey} \\
  {\small \tt osert@pamukkale.edu.tr}}
\vskip 1cm
\date{ }
\maketitle

 \thispagestyle{empty}

\begin{abstract}

We develop a symmetric teleparallel gravity model in a space-time
with only the non-metricity is nonzero, in terms of a Lagrangian
quadratic in the non-metricity tensor. We present a detailed
discussion  of the variations that may be used for any
gravitational formulation. We seek Schwarzschild-type solutions
because of its observational significance and obtain a class of
solutions that includes Schwarzschild-type, Schwarzschild-de
Sitter-type and Reissner-Nordstr\"{o}m-type solutions for certain
values of the parameters. We also discuss the physical relevance
of these solutions.

\end{abstract}
\end{titlepage}

\section{\large Introduction}

Although Einstein's (pseudo-)Riemannian formulation of
gravitation, the so-called general relativity (GR), is elegant and
successful, there are other hints to study non-Riemannian
generalizations \cite{adak2003}. In GR, only the dynamical field
is the metric or the co-frame and the corresponding field strength
is the curvature ${R^a}_b$ written in terms of the Levi-Civita
connection. In non-Riemannian gravity models the geometry is
enriched by adding torsion $T^a$ and non-metricity ${Q^a}_b$ that
modify the Levi-Civita connection ${\omega^a}_b$ of the Riemannian
space-time (see \cite{adak2003}-\cite{hehl1995} and references
therein). The orthonormal components of the metric $\eta_{ab}$,
orthonormal co-frame $1-$forms $e^a$ and the full connection
$1-$fotms ${\Lambda^a}_b$ are considered as gauge potentials in
the gauge approach to gravity. The corresponding field strengths
are defined by the covariant exterior derivatives of the
potentials; namely the non-metricity, torsion and curvature,
respectively. The most general non-Riemannian gravity models ($Q
\neq 0$, $T \neq 0$, $R \neq 0$) can be found in the literature
\cite{adak2003}-\cite{adak2004}. An important subclass of these
models is provided by Einstein-Cartan-Dirac theory written in a
Riemann-Cartan space-time in which a spinor field is the source of
space-time torsion \cite{dereli1982} and the torsion may
contribute to neutrino oscillations \cite{adak2001}. We emphasize
that the connection is metric compatible ($Q=0$) in such theories.
In certain modifications of this model, curvature together with
non-metricity is also constrained to zero, but torsion is left
nonzero. These theories are called teleparallel gravity models
(see \cite{hayashi1967}-\cite{arcos2004} and references therein).

Another approach to non-Riemannian gravity is due to H. Weyl who
used only the trace of the non-metricity 1-form, i.e. Weyl 1-form,
and in an unsuccessful attempt interpreted that as the
electromagnetic potential 1-form. In addition to this pioneering
work, there have been other theoretical models in space-times with
$Q \neq 0$, $R \neq 0$, $T=0$ (\cite{dereli1994}-\cite{hehl2005}
and references therein). On the other hand, gravity models in the
space-times with $Q \neq 0$, $T=0$, $R=0$, that we are going to
call symmetric teleparallel gravity (STPG), have not been studied
that much \cite{nester1999}-\cite{adak2005}. This paper aims to
fill this gap. Due to the fact that curvature and torsion vanish,
it is usually asserted that this model is a gravitational field
theory that is closest possible to flat space-time. This is an
interesting aspect of STPG, that deserves further analysis.

After recalling some mathematical preliminaries in
Sec.\ref{matpre}, we derive the field equations of STPG from the
most general quadratic non-metricity Lagrangian in
Sec.\ref{model}. Mathematical details of the variations are given
in an appendix.  Then in Sec.\ref{solution} we write down the
nontrivial dynamical gravitational field equations for a
spherically symmetric background metric and find a class of
solutions that includes a Schwarzschild-type solution, in which we
identify the integration constant with a spherically symmetric
mass located at the coordinate center; Schwarzschild-de
Sitter-type, in which we we identify one of the integration
constants with the cosmological constant, and a
Reissner-Nordstr\"{o}m-type solution. The last section is reserved
for the conclusion. We speculate that dark matter may couple to a
new kind of gravitational charge and this speculative charge may
interact gravitationally only through non-metricity. Here we
adhere to the following conventions: Indices denoted by Greek
letters $\alpha$, $\beta , \;\; \cdots =
\hat{0},\hat{1},\hat{2},\hat{3}$ are holonomic or coordinate
indices, $a$, $b, \;\; \cdots = 0,1,2,3$ are anholonomic or frame
indices. We abbreviate the wedge products of orthonormal co-frames
$e^{ab \cdots} = e^a \wedge e^b \wedge \cdots$ and denote
symmetric and anti-symmetric indices as $(ab)=1/2 (a+b)$ and $[ab]
=1/2 (a-b)$.

\section{\large Mathematical preliminaries}\label{matpre}

Spacetime is denoted by the triple $ \{M,g,\nabla \} $ where M is
a 4-dimensional differentiable manifold equipped with a Lorentzian
metric $ g $ which is a second rank, covariant, symmetric,
non-degenerate tensor and $ \nabla $ is a linear connection which
defines parallel transport of vectors (or more generally tensors
and spinors). In terms of an orthonormal basis $\{ X_a \}$, the
metric can be written as
 \ba
      g = \eta_{ab}e^a \otimes e^b \;\;\; , \;\; a,b,\cdots = 0,1,2,3
        \label{metric}
 \ea
where $\eta_{ab}=(-,+,+,+)$ is the Minkowski metric and $\{ e^a
\}$ is the orthonormal co-frame. The local orthonormal frame $\{
X_a \}$ is dual to the co-frame $\{e^a \}$;
 \ba
     e^b(X_a)= \imath_a e^b = \delta^b_a
 \ea
where $\imath_{X_a} \equiv \imath_a$ are the interior product
operators. The space-time orientation is set by the choice
$\epsilon_{0123}=+1 $ or $*1 =e^{0123}$ where $*$ denotes the
Hodge dual map. In addition, the connection is specified by a set
of connection 1-forms $\{ {\Lambda^a}_b \}$. In the gauge approach
to gravity, $\eta_{ab} , \quad e^a , \quad {\Lambda^a}_b$ are
interpreted as the generalized gauge potentials, while the
corresponding field strengths; the non-metricity 1-forms, torsion
2-forms and curvature 2-forms are defined through the Cartan
structure equations.

  \subsection{Nonmetricity : \hskip 0.3cm $\mathbf{{Q^a}_b}$}
 \ba
   2Q_{ab} := -D\eta_{ab} = \Lambda_{ab} +\Lambda_{ba} \label{nonmet}
 \ea
where $D$ denotes the covariant exterior derivative. When
$Q_{ab}=0$, it is said that the connection is {\it metric
compatible}. In this case $\Lambda_{ab} = - \Lambda_{ba}$.
Geometrically the non-metricity tensor measures the deformation of
length and angle standards during parallel transport. Technically
speaking it is a measure of compatibility of the affine connection
with the metric. The scalar product of vectors is, in general, not
preserved during parallel transport due to the appearance of
non-metricity. As a result if we parallel transport a vector along
a closed curve, the length of the final vector will be different
from that of the initial vector.

  \subsection{Torsion : \hskip 0.3cm $\mathbf{T^a}$}
 \ba
   T^a := De^a = de^a + {\Lambda^a}_b \wedge e^b  \label{torsion}
 \ea
where $d$ denotes exterior derivative. If the connection is not
only metric compatible, but the torsion is also zero, i.e, both $
Q_{ab} = 0 $ and $ T^a = 0 $, it is said that the connection is
{\it Levi-Civita}. Geometrically it relates to the translational
group. If a vector (say $ A=X_a A^a $) is parallel transported
around a loop by means of the linear connection $ {\Lambda^a}_b $,
a transformation of $ A $ is induced, the linear piece of which is
determined by the curvature and the translational piece by the
torsion. To see its translational part let us consider parallel
transport. Suppose that a vector $ A = A^\mu \partial_\mu $ is
parallelly transported according to the following prescription:
First parallel transport the vector along $ dx^\mu $ and then
along $ d x^\nu $. Compare that with repeating the trip in reverse
order, that is, first transport the vector along  $ d x^\nu $ and
then along $ dx^\mu $. In flat space-time this series of actions
sketches out a parallelogram. What happens in curved space? As we
will see later, with zero torsion one still obtains a
parallelogram, but of course the vector ends up rotated. With
non-zero torsion something else happens; the two different routes
do not bring the vector to the same point. The total difference
between the initial and the final points of the vector, $ \Delta
A^a $, is given by
 \ba
  \Delta A^a \simeq \frac{1}{2} \int_S{{T^a}_{\mu \nu} dx^\mu \wedge dx^\nu}
 \ea
where $ S $ is the surface enclosed by the loop. It is sometimes
said that closed parallelograms do not exist in space-time with
torsion.

 \subsection{ Curvature : \hskip 0.3cm $\mathbf{{R^a}_b}$}
 \ba
  {R^a}_b := D{\Lambda^a}_b :=
     d{\Lambda^a}_b +{\Lambda^a}_c \wedge {\Lambda^c}_b \; .
     \label{curva}
 \ea
Geometrically it relates to the linear group. Now let us see the
effect of curvature on vectors after being parallel transported
along a closed loop. If the vector does not undergo a rotation,
then space is flat. Conversely, if the vector is rotated, then
space is curved. Performing the parallel transport of a vector $ A
= X_a A^a $ around a closed small path one obtains the following
transformation
 \ba
 \Delta A^a \simeq \frac{1}{2}\int_S{{R^a}_{b \mu \nu } A^b dx^\mu \wedge dx^\nu }
 \ea
where $ S $ is the surface of the loop. Space-time is curved if
this tensor is not zero, and the space-time is flat if this tensor
is zero. Physically space-time curvature is a fundamental tensor
in gravitation, in particular in Einstein's general relativity.

These field strengths satisfy the Bianchi identities
 \ba
     DQ_{ab} &=& \frac{1}{2}(R_{ab} +R_{ba}), \label{bian1}\\
     DT^a    &=& {R^a}_b \wedge e^b , \label{bian2}\\
     D{R^a}_b &=& 0 . \label{bian3}
 \ea
We also need the identities \footnote{Since
$Q^{ab}=\frac{1}{2}D\eta^{ab} \neq 0$ we pay special attention in
lowering and raising an index in front of the covariant exterior
derivative.}
 \ba
    D*e_a &=& - Q \wedge *e_a + T^b \wedge *e_{ab} \\
    D*e_{ab} &=& - Q \wedge *e_{ab} + T^c \wedge *e_{abc} \\
    D*e_{abc} &=& - Q \wedge *e_{abc} + T^d \wedge *e_{abcd} \\
    D*e_{abcd} &=& - Q \wedge *e_{abcd}
 \ea
where $Q={\Lambda^a}_a={Q^a}_a$ is the Weyl 1-form. The linear
connection 1-forms can be decomposed uniquely as follows
\cite{dereli1987}-\cite{hehl1995}:
 \ba
  {\Lambda^a}_b = {\omega^a}_b + {K^a}_b
          + {q^a}_b + {Q^a}_b \;  \label{connec}
 \ea
where $ {\omega^a}_b $ are the  Levi-Civita connection 1-forms
 \ba
     de^a + {\omega^a}_b \wedge e^b = 0 \; , \label{levi}
 \ea
$ {K^a}_b $ are the contortion 1-forms
 \ba
     {K^a}_b \wedge e^b = T^a \; , \label{contor}
 \ea
and $ {q^a}_b $ are the anti-symmetric tensor 1-forms
 \ba
     q_{ab} = -(\imath_a Q_{bc}) \wedge e^c
        + (\imath_b Q_{ac}) \wedge e^c \, . \label{antisy}
 \ea
In the above  decomposition the symmetric part
 \ba
    \Lambda_{(ab)} = Q_{ab} \label{symm}
 \ea
while the anti-symmetric part
 \ba
  \Lambda_{[ab]} = \omega_{ab} + K_{ab} + q_{ab} \; . \label{asymm}
 \ea
It is cumbersome to take into account all components of
non-metricity in gravitational models. Therefore we will be
content with dealing  only with certain irreducible parts of it to
gain physical insight. The irreducible decompositions of
non-metricity invariant under the Lorentz group are summarily
given below \cite{hehl1995}. The non-metricity 1-forms $ Q_{ab} $
can be split into their trace-free $ \overline{Q}_{ab} $ and the
trace parts as
 \ba
   Q_{ab} = \overline{Q}_{ab} + \frac{1}{4} \eta_{ab}Q          \label{118}
 \ea
where the Weyl 1-form $Q={Q^a}_a$ and $ \eta^{ab}\overline{Q}_{ab}
= 0 $. Let us define
 \ba
    \Lambda_b &:=& \imath_a { \overline{Q}^a}_b \;\; , \;\;\;\;\;\;\;\;
                   \Lambda := \Lambda_a e^a,    \nonumber  \\
    \Theta_b &:=& {}*(\overline{Q}_{ab} \wedge e^a) \;\; , \;\;\;
    \Theta := e^b \wedge \Theta_b \;\; , \;\;\;
    \Omega_a := \Theta_a -\frac{1}{3}\imath_a\Theta    \label{119}
 \ea
as to use them in the decomposition of $ Q_{ab} $ as
 \ba
     Q_{ab} = { }^{(1)}Q_{ab} + { }^{(2)}Q_{ab} +
            { }^{(3)}Q_{ab} + { }^{(4)}Q_{ab}             \label{120}
 \ea
where
 \ba
    { }^{(2)}Q_{ab} &=& \frac{1}{3} {}*(e_a \wedge \Omega_b +e_b
    \wedge \Omega_a) ,  \\
    { }^{(3)}Q_{ab} &=& \frac{2}{9}( \Lambda_a e_b +\Lambda_b e_a
                      -\frac{1}{2} \eta_{ab} \Lambda ) , \\
    { }^{(4)}Q_{ab} &=&  \frac{1}{4} \eta_{ab} Q   , \\
    { }^{(1)}Q_{ab}  &=& Q_{ab}- { }^{(2)}Q_{ab}
                        - { }^{(3)}Q_{ab} - { }^{(4)}Q_{ab} \;.
 \ea
We have
 \ba
   \eta_{ab} { }^{(1)}Q^{ab} = \eta_{ab} { }^{(2)}Q^{ab}
     =\eta_{ab} { }^{(3)}Q^{ab} = 0 \; ,  \nonumber \\
 \imath_a { }^{(1)}Q^{ab} = \imath_a { }^{(2)}Q^{ab} =0 \;, \nonumber \\
  e_a \wedge { }^{(1)}Q^{ab} =0 \; ,  \nonumber \\
  \imath_{(a}{ }^{(2)}Q_{bc)} =0 \; .
 \ea
Thus the components are orthogonal in the following sense
 \ba
    { }^{(i)}Q^{ab} \wedge \; * { }^{(j)}Q_{ab} = \delta^{ij}
    N_{ij} \quad \quad (\mbox{\small no summation over {\it ij}})
 \ea
where $\delta^{ij}$ is the Kronecker symbol and $N_{ij}$ is a
$4$-form. Then
 \ba
 { }^{(1)}Q^{ab} \wedge \; * { }^{(1)}Q_{ab} &=& Q^{ab} \wedge \;* Q_{ab}
 - { }^{(2)}Q^{ab} \wedge \; * { }^{(2)}Q_{ab}
 - { }^{(3)}Q^{ab} \wedge \; * { }^{(3)}Q_{ab}  \nonumber \\
 & & \quad \quad \quad \quad - { }^{(4)}Q^{ab} \wedge \; * { }^{(4)}Q_{ab} \; , \label{Q1Q1} \\
 { }^{(2)}Q^{ab} \wedge \; * { }^{(2)}Q_{ab} &=& \frac{2}{3} (Q_{ac} \wedge e^a) \wedge * (Q^{bc} \wedge e_b)
 - \frac{2}{9} (\imath^a Q_{ac}) (\imath_b Q^{bc}) *1 - \frac{2}{9} Q \wedge * Q \nonumber \\
 & & \quad \quad \quad \quad + \frac{4}{9} (\imath_a Q) (\imath_b Q^{ab}) *1 \; , \\
 { }^{(3)}Q^{ab} \wedge \; * { }^{(3)}Q_{ab} &=& \frac{4}{9} (\imath^a
 Q_{ac}) (\imath_b Q^{bc}) *1 + \frac{1}{36} Q \wedge * Q
 - \frac{2}{9} (\imath_a Q) (\imath_b Q^{ab}) *1 \; , \\
 { }^{(4)}Q^{ab} \wedge \; * { }^{(4)}Q_{ab} &=& \frac{1}{4}   Q \wedge * Q \;
 . \label{Q4Q4}
 \ea

\section{\large Symmetric teleparallel gravity }\label{model}

We formulate STPG in terms of a Lagrangian 4-form
 \ba
   \mathcal{L} = L + \lambda_a \wedge T^a + {R^a}_b \wedge  {\rho_a}^b \label{lagranj1}
 \ea
where ${\rho_a}^b$ and $\lambda_a$ are the Lagrange multiplier
2-forms giving the constraints
 \ba
   {R^a}_b = 0 \quad , \quad \quad T^a = 0 \; .
 \ea
$\mathcal{L}$ changes by a closed form under the transformations
 \ba
      \lambda_a & \rightarrow & \lambda_a + D\mu_a \; , \label{lama}\\
   {\rho_a}^b   & \rightarrow &{\rho_a}^b + D{\xi_a}^b - \mu_a \wedge
   e^b \label{rhoab}
 \ea
of the Lagrange multiplier fields. Here $\mu_a$ and ${\xi_a}^b$
are arbitrary 1-forms. To show this invariance we use the Bianchi
identities (\ref{bian1})-(\ref{bian3}) and discard exact forms.
Consequently the field equations derived from the Lagrangian
$4-$form (\ref{lagranj1}) will determine the Lagrange multipliers
only up to above transformations. The gravitational field
equations are derived from (\ref{lagranj1}) by independent
variations with respect to the connection $ \{ {\Lambda^a}_b \} $
and the ortohonormal co-frame $\{ e^a \}$ 1-forms, respectively:
 \ba
     \lambda_a \wedge e^b + D{\rho_a}^b = - {\Sigma_a}^b  \; , \label{dLambda1}  \\
  D\lambda_a = - \tau_a  \label{dcoframe1}
 \ea
where ${\Sigma_a}^b = \frac{\partial L}{\partial {\Lambda^a}_b}$
and $\tau_a = \frac{\partial L}{\partial e^a}$. In principle the
first field equation (\ref{dLambda1}) is used to solve for the
Lagrange multipliers $\lambda_a$ and ${\rho_a}^b$ and the second
field equation (\ref{dcoframe1}) governs the dynamics of the
gravitational fields. Here the first equation, however, has $64$
and the second one has $16$ independent components, thus giving
the total number of independent equations $80$. On the other hand,
there are totally $120$ unknowns: $24$ for $\lambda_a$ plus $96$
for ${\rho_a}^b$. But we note that the left-hand side of
(\ref{dLambda1}) is invariant under the transformations
(\ref{lama})-(\ref{rhoab}) and consequently it is sufficient to
determine the gauge invariant piece of the Lagrange multipliers,
namely $\lambda_a \wedge e^b + D{\rho_a}^b$, in terms of
${\Sigma_a}^b$. It is important to notice $D \lambda_a$ rather
than the Lagrange multipliers themselves couple to the second
field equations (\ref{dcoframe1}). As a result we must calculate
$D\lambda_a$ directly and we can manage that by taking the set of
covariant exterior derivative of (\ref{dLambda1}):
 \ba
     D\lambda_a \wedge e^b = - D {\Sigma_a}^b \; .  \label{Dlambda}
 \ea
Here we used the constraints
 \ba
     De^b &=& T^b =0 ,\\
 D^2{\rho_a}^b &=& D(D{\rho_a}^b) = {R^b}_c \wedge {\rho_a}^c - {R^c}_a \wedge
 {\rho_c}^b =0 \label{D2Sab}
 \ea
where the covariant exterior derivative of a $(1,1)$-type tensor
is
 \ba
    D{\rho_a}^b = d {\rho_a}^b + {\Lambda^b}_c \wedge {\rho_a}^c
         - {\Lambda^c}_a \wedge {\rho_c}^b \; .
 \ea
The result (\ref{Dlambda}) is unique because $D\lambda_a
\rightarrow D\lambda_a$ under (\ref{lama}). One can consult
Ref.\cite{vasilic2000} for further discussions on gauge symmetries
of Lagrange multipliers. Thus we arrive at the field equation
 \ba
     D{\Sigma_a}^b -  \tau_a \wedge e^b =0 \; . \label{fieldeqn}
 \ea

Now we write down the following Lagrangian 4-form which is the
most general quadratic expression in the non-metricity tensor
\cite{obukhov1997}:
 \ba
   L = \frac{1}{2\kappa} \left[ k_0 {R^a}_b \wedge *{e_a}^b
    + \sum_{I=1}^4 k_I { }^{(I)}Q_{ab} \wedge * { }^{(I)}Q^{ab}
    + k_5 \left( { }^{(3)}Q_{ab} \wedge e^b \right) \wedge * \left( { }^{(4)}Q^{ac} \wedge e_c
    \right) \right] \; . \label{Lagrange}
   \ea
Here $k_0, k_1, k_2, k_3, k_4, k_5$ are dimensionless coupling
constants and $\kappa = \frac{8\pi G}{c^3}$, with $G$ the Newton's
gravitational constant. Inserting (\ref{Q1Q1})-(\ref{Q4Q4}) into
(\ref{Lagrange}) we find
 \ba
   L = \frac{1}{2\kappa} \left[ k_0 {R^a}_b \wedge *{e_a}^b
    + c_1 Q_{ab} \wedge * Q^{ab}
      +c_2 (Q_{ac} \wedge e^a) \wedge * (Q^{bc} \wedge e_b)  \right. \nonumber \\
      \left. + c_3 (\imath_a Q^{ac}) (\imath^b Q_{bc}) *1
      +c_4 Q \wedge * Q
       +c_5 (\imath_a Q) (\imath_b Q^{ab}) *1 \right] \label{lagranj}
   \ea
where the new coefficients are the following combinations of the
original coupling constants:
 \ba
 c_1 &=& k_1 ,\nonumber \\
 c_2 &=& - \frac{2}{3} k_1 + \frac{2}{3} k_2 \; ,\nonumber \\
 c_3 &=& - \frac{2}{9} k_1 - \frac{2}{9} k_2 + \frac{4}{9} k_3 \;, \nonumber \\
 c_4 &=& - \frac{1}{18} k_1 - \frac{2}{9} k_2 + \frac{1}{36} k_3 + \frac{1}{4} k_4 + \frac{1}{16} k_5 \;, \nonumber \\
 c_5 &=& - \frac{2}{9} k_1 + \frac{4}{9} k_2 - \frac{2}{9} k_3 - \frac{1}{4} k_5 \; .
 \ea
When we  use the results derived in the appendix by keeping in
mind $Q^{ab} = \Lambda^{(ab)}$, together with (\ref{fieldeqn}) we
obtain the variational field equations from (\ref{lagranj})
 \ba
   {\Sigma_a}^b = \sum_{i=0}^{5} c_i \; { }^i{\Sigma_a}^b \quad ,
   \quad \quad \quad \quad
  \tau_a = \sum_{i=0}^{5} c_i \; { }^i\tau_a
 \ea
where
 \ba
   { }^0{\Sigma_a}^b  &=& 2Q^{bc} \wedge *e_{ac} - Q \wedge *{e_a}^b + T_c \wedge *{e_a}^{bc} \\
     { }^1{\Sigma_a}^b  &=& *({Q_a}^b + {Q^b}_a) \\
  { }^2{\Sigma_a}^b  &=& e_a \wedge *(Q^{bc} \wedge e_c) + e^b \wedge *(Q_{ac} \wedge e^c) \\
   { }^3{\Sigma_a}^b  &=& \imath^c Q_{ac} *e^b + \imath_c Q^{bc} *e_a \\
    { }^4{\Sigma_a}^b  &=& 2 \delta^b_a { }*Q \\
    { }^5{\Sigma_a}^b  &=& \frac{1}{2} (\imath^b Q) *e_a +  \frac{1}{2} (\imath_a Q) *e^b
                          + \delta^b_a  (\imath_c Q^{cd}) *e_d \\
    { }^0\tau_a  &=&  {R^b}_c \wedge *{e_{ab}}^c \\
    { }^1\tau_a  &=&  - (\imath_a Q^{bc}) \wedge *Q_{bc}
                              - Q^{bc} \wedge (\imath_a *Q_{bc}) \\
    { }^2\tau_a  &=& -Q_{ab} \wedge *(Q^{bc} \wedge e_c)
                            - (\imath_a Q^{cd}) e_c \wedge *(Q_{bd} \wedge e^b )
                             + (Q_{bd} \wedge e^b) \wedge *(Q^{cd} \wedge e_{ca}) \\
    { }^3\tau_a  &=& -2 (\imath_a Q^{bd}) (\imath^c Q_{cd}) *e_b
                             + (\imath_b Q^{bd}) (\imath^c Q_{cd}) *e_a \\
    { }^4\tau_a  &=&  -(\imath_a Q) *Q - Q \wedge (\imath_a *Q) \\
    { }^5\tau_a  &=&   (\imath_b Q) (\imath_c Q^{bc}) *e_a
                            - (\imath_a Q) (\imath_c Q^{bc}) *e_b
                            - (\imath_b Q) (\imath_a Q^{bc}) *e_c \; .
 \ea
Since ${ }^0\tau_a  =  {R^b}_c \wedge *{e_{ab}}^c = 0$ and $D {
}^0{\Sigma_a}^b = D^2 *{e_a}^b \sim {R_a}^b =0 $ we drop the
Einstein-Hilbert term: $k_0 =0$. The case $k_0 \neq 0$ and
others$=0$ is discussed in \cite{adak2005}.

\subsection{Spherical symmetric solution to the
model}\label{solution}

We now proceed the attempt for finding a solution to the STPG
model. We naturally start dealing with the case of spherical
symmetry for realistic simplicity. As usual in the study of exact
solutions one has to choose the convenient local coordinates and
make corresponding ansatz for the dynamical fields. We try to find
a spherically symmetric solution with the line element
 \ba
     g=-F^2 dt^2 + G^2dr^2 + r^2d\theta^2 +r^2\sin^2\theta
     d\varphi^2
 \ea
where $F=F(r)$ and $G=G(r)$. A convenient choice for a tetrad
reads
 \ba
      e^0 = Fdt , \quad e^1= Gdr , \quad e^2= rd\theta ,
      \quad e^3 = r\sin\theta d\varphi \; . \label{coframe}
 \ea
In addition to orthonormal co-frame choice, for the non-Riemannian
connection we choose
 \ba
     \Lambda_{12} &=& -\Lambda_{21}= - \frac{1}{r}e^2 , \quad
     \Lambda_{13}=-\Lambda_{31}= -  \frac{1}{r}e^3  , \quad
     \Lambda_{23}=-\Lambda_{32}= - \frac{\cot\theta}{r}e^3 , \nonumber \\
     \Lambda_{00} &=& \frac{F'}{FG}e^1 , \quad
     \Lambda_{11} = \Lambda_{22} = \Lambda_{33} = \frac{1}{r}(1-\frac{1}{G})e^1 , \quad
     \mbox{others}=0   \; . \label{connect}
 \ea
Here prime denotes the derivative with respect to $r$. This gauge
configuration satisfies the constraint equations ${R^a}_b(\Lambda
)=0 \quad , \quad T^a(\Lambda )=0$.
Equations(\ref{coframe})-(\ref{connect}) read explicitly
 \ba
    \omega_{01} &=& -\frac{F'}{FG}e^0 , \quad \omega_{12}=- \frac{1}{rG} e^2 , \quad
     \omega_{13}=- \frac{1}{rG} e^3 , \quad
     \omega_{23}= -\frac{\cot\theta}{r} e^3 , \nonumber \\
  Q_{00} &=&  \frac{F'}{FG}e^1 , \quad Q_{11} = Q_{22} = Q_{33} = \frac{1}{r}(1-\frac{1}{G})e^1 , \quad
     \mbox{others}=0  \label{oQq} \; .
 \ea
Under this configuration the only nontrivial field equation comes
from the trace of (\ref{fieldeqn}):
 \ba
     d {\Sigma^a}_a + e^a \wedge \tau_a =0 \; . \label{65}
 \ea
Symmetric and antisymmetric parts of the field equation give
trivially zero.  From (\ref{65}) we obtain
 \ba
     - \frac{\ell_1}{G} \left( \frac{F'}{FG} \right)'
     - \frac{\ell_2}{G} \left( \frac{1-G}{rG} \right)'
     + \ell_3 \left( \frac{F'}{FG} \right)^2
     + \ell_4 \left( \frac{F'}{FG} \right) \left( \frac{1-G}{rG} \right)
     - \frac{2\ell_1}{rG} \left( \frac{F'}{FG} \right) \nonumber \\
     - \frac{2\ell_2}{rG} \left( \frac{1-G}{rG} \right)
     + \ell_5 \left( \frac{1-G}{rG} \right)^2 = 0
 \ea
where
 \ba
 \ell_1 &=& 2c_1 + 2c_2 + 8c_4 + c_5 ,\\
 \ell_2 &=& 6c_1 + 4c_2 + 2c_3 + 24c_4 + 7c_5 ,\\
 \ell_3 &=& - 6c_4 - c_5 ,\\
 \ell_4 &=& -6c_1 - 4c_2 - 2c_3 - 12c_4 - 5c_5 ,\\
 \ell_5 &=& 6c_1 + 4c_2 + 2c_3 + 18c_4 + 6c_5 \; .
 \ea
Due to the observational success of Schwarzschild solution of
general relativity, we investigate solutions with $ G=1/F $. Then
 \ba
  - \ell_1 FF'' + \ell_3 (F')^2 - (2\ell_1 + \ell_2 - \ell_4) \frac{FF'}{r}
     + (\ell_5 - \ell_2) \frac{F^2}{r^2} - \ell_4 \frac{F'}{r} + ( \ell_2 - 2\ell_5) \frac{F}{r^2}
     + \ell_5 \frac{1}{r^2}= 0 \; . \label{srn-equation}
 \ea
We can not find an analytical exact solution of this nonlinear
second order differential equation. Therefore, we treat the linear
sector of the equation
 \ba
     r^2 \left(F^2\right)'' + (2+\frac{\ell_2}{\ell_1}) r
     \left(F^2\right)' + \frac{\ell_2}{\ell_1} F^2 = \frac{\ell_2}{\ell_1}
 \ea
by choosing our parameters as follows;
 \ba
     \ell_3 = -\ell_1  \quad , \quad  \ell_4 =0  \quad , \quad \ell_5 = \frac{\ell_2}{2} \; .
 \ea
Here some special cases deserve attention.

\begin{enumerate}
    \item For $\ell_2 =\ell_1 $, we obtain the solution
                  \ba
               F^2 = 1 + \frac{C_1}{r} + D_1 \frac{\ln{r}}{r}
                 \ea
            which is asymptotically flat; $\lim_{r \rightarrow \infty} F = 1$.
            Here $C_1$ and $D_1$ are integration constants.
    \item For $\ell_2 \neq \ell_1 $, the solution is found as
                \ba
              F^2 = 1 + \frac{C_2}{r} + \frac{D_2}{r^{\ell_2 / \ell_1}}
                \ea
           where $C_2$ and $D_2$ are integration constants.
\begin{enumerate}
    \item  For $\ell_2 =0$, we obtain a Schwarzschild-type solution with
           $D_2=0$ for asymptotically flatness and we identify the other
           constant with a spherically symmetric mass centered at the origin; $C_2 = -2M$.
    \item  For $\ell_2 =-2\ell_1$, we obtain a Schwarzschild-de Sitter-type solution.
           We again identify $C_2$ with mass $C_2 = -2M$ and $D_2$ with
           cosmological constant \footnote{Current estimates suggest $|
           \Lambda | < 10^{-54}$ cm$^{-2}$, and this makes $\Lambda$ quite
           negligible in all noncosmological situations.}
           $D_2=-\frac{1}{3}\Lambda$. The $\Lambda$ term corresponds to a
           repulsive central force of magnitude $\frac{1}{3}\Lambda r$, which
           is independent of the central mass.
    \item  For $\ell_2 = 2\ell_1$, we obtain a Reissner-Nordstr\"{o}m-type solution.
           We again identify $C_2$ with mass $C_2 = -2M$ while $D_2$ with a new
           kind of gravitational charge. We hope that besides ordinary matter
           that interacts gravitationally through its mass, the dark matter
           in the Universe may interact gravitationally through both its mass
           and this new gravitational charge.
\end{enumerate}
\end{enumerate}

\section{Conclusion}

Einstein's GR provides a (pseudo-)Riemannian formulation of
gravitation in the absence of matter. The field equations are
obtained as the local extremum of the Einstein-Hilbert action
integral under metric variations. The integrand of this action is
simply the curvature scalar of the space-time curvature written in
terms of the Levi-Civita connection. Such a connection is
torsion-free and metric compatible and provides a useful reference
connection because it depends entirely on the metric. Einstein's
GR is very successful for describing the large scale structure of
gravitational phenomena. But in spite of that, supergravity and
superstring theories have suggested a more general geometry with
torsion and non-metricity. At the level of effective theories
there are hints that a non-Riemannian geometry may offer a more
economical and more elegant description of gravitational
interactions. Although theories in which non-Riemannian
geometrical fields are dynamical in the absence of matter are more
difficult to interpret, they may play an important role in certain
astrophysical contexts. In this context here we have studied the
Lagrange formulation of the general symmetric teleparallel gravity
model, a theory in which only non-metricity is nonvanishing. We
have not seen a similar analysis of STPG in the literature. We
consider the full 5-parameter symmetric teleparallel Lagrangian
without {\it a priori} restricting the coupling constants $c_1,
c_2, c_3, c_4, c_5$. The main motivation for this was to determine
the place and significance of STPG in the framework of
non-Riemannian models. We found that the field equations accept a
Schwarzschild-type solution for some special values of the
coupling constants. For some other set of values of the parameters
we found a Schwarzscild-de Sitter-type solution. A further
different set of values of the coupling constants yields a
Reissner-Nordstr\"{o}m-type solution. Consequently, we suggest
that in addition to ordinary matter that interacts gravitationally
through its mass, the dark matter in the Universe may interact
gravitationally through both its mass and a new kind of
gravitational charge \cite{tucker1998}-\cite{dereli2000}. The
latter coupling is analogous to the coupling of electric charge to
the photon where the analogue of the Maxwell field here is the
non-metricity field strength. Such novel gravitational
interactions may have a significant influence on the structure of
black holes. For example, we may speculate that this unknown
gravitational charge may have repulsive nature and thus black
holes may not need to be close in one end. Possible matter
couplings to the model and some physical events such as Berry
phase and neutrino oscillations in STPG will be discussed
elsewhere.

\section*{Acknowledgment}

We are grateful to  Prof. Dr. Tekin DEREL\.I for fruitful
discussions.

 \appendix

\section{Calculus of variation}\label{variation}

Let $M$ be a $n$-dimensional manifold and $\alpha , \beta \in
\Lambda^p(M)$ where $\Lambda^p(M)$ denotes any $p$-form on $M$.
Then we would like to find the extremum of the following action
integral
 \ba
   I [\alpha , \beta , e^a] = \int_M {\alpha \wedge * \beta}
 \ea
by varying it with respect to the variables; $\alpha$, $\beta$ and
$e^a$:
 \ba
   \delta I = \int_M {\delta \alpha \wedge * \beta + \alpha \wedge \delta *
   \beta} \label{deltaI}
 \ea
The second term on the right hand side needs a detailed
calculation, because it contains the Hodge star.
 \ba
 \alpha \wedge \delta * \beta &=& \alpha \wedge \delta (\frac{1}{p!} \beta_{i_1 \cdots i_p}
                                    * e^{i_1 \cdots i_p}) \nonumber \\
                      &=& \alpha \wedge \frac{1}{p!} ( \delta \beta_{i_1 \cdots i_p})
                          * e^{i_1 \cdots i_p} + \alpha \wedge \frac{1}{p!} \beta_{i_1 \cdots i_p}
                           \delta * e^{i_1 \cdots i_p}
 \ea
First use the identity in the first term $ \theta \wedge * \gamma
= \gamma \wedge * \theta $ where $\theta \; , \; \gamma \in
\Lambda^p(M)$
 \ba
 \alpha \wedge \delta * \beta = \frac{1}{p!} (\delta \beta_{i_1 \cdots i_p})
    e^{i_1 \cdots i_p} \wedge * \alpha + \alpha \wedge \frac{1}{p!} \beta_{i_1 \cdots i_p}
                           \delta * e^{i_1 \cdots i_p}
                           \label{alphadelta*beta}
 \ea
and using
 \ba
 \delta \beta &=& \delta ( \frac{1}{p!} \beta_{i_1 \cdots i_p} e^{i_1 \cdots i_p}) \nonumber \\
     &=& \frac{1}{p!} (\delta \beta_{i_1 \cdots i_p}) e^{i_1 \cdots i_p}
     + (\delta e^{i_1}) \wedge \frac{1}{(p-1)!} \beta_{i_1 \cdots i_p} e^{i_2 \cdots
     i_p} \nonumber \\
  &=& \frac{1}{p!} (\delta \beta_{i_1 \cdots i_p}) e^{i_1 \cdots i_p}
     + (\delta e^{i_1}) \wedge (\imath_{i_1} \beta) \\
 \frac{1}{p!} (\delta \beta_{i_1 \cdots i_p}) e^{i_1 \cdots i_p}
 &=& \delta \beta - (\delta e^a) \wedge (\imath_a \beta)
 \ea
and the equality
  \ba
  \frac{1}{p!} \beta_{i_1 \cdots i_p} \delta * e^{i_1 \cdots i_p}
  &=& \frac{1}{p!} \beta_{i_1 \cdots i_p} \delta \left[ \frac{1}{(n-p)!}
  {\epsilon^{i_1 \cdots i_p}}_{i_{p+1} \cdots i_n} e^{i_{p+1} \cdots
  i_n} \right] \nonumber \\
  &=& (\delta e^{i_{p+1}}) \wedge    \frac{1}{p!(n-p-1)!}
  {\epsilon^{i_1 \cdots i_p}}_{i_{p+1} \cdots i_n} \beta_{i_1 \cdots i_p} e^{i_{p+2} \cdots
  i_n}  \nonumber \\
  &=& (\delta e^a) \wedge (\imath_a * \beta)
 \ea
in (\ref{alphadelta*beta}) and then substituting the result into
(\ref{deltaI}) we obtain
  \ba
   \delta I &=& \delta \int_M \alpha \wedge * \beta \nonumber \\
   &=& \int_M {\delta \alpha \wedge * \beta + \delta \beta \wedge * \alpha
   -\delta e^a \wedge [(\imath_a \beta) \wedge * \alpha - (-1)^p \alpha \wedge (\imath_a *
   \beta)]} \label{deltaI2}
 \ea
where $\alpha \; , \; \beta \in \Lambda^p(M)$.

 \begin{center}
 \begin{tabular}{|c|}
   \hline
     {\it \large Special Case:1} \hskip 1cm $\alpha = \beta = F = dA$\\
   \hline
 \end{tabular}
 \end{center}

We encounter these kinds of Lagrangians in electromagnetic theory
and teleparallel gravity models. In this case (\ref{deltaI2})
becomes
  \ba
   \delta I = \int_M { (\delta dA) \wedge (2* dA)
   - \delta e^a \wedge [(\imath_a F) \wedge * F - (-1)^p F \wedge (\imath_a
   *F)]} \label{85}
 \ea
where $A \in \Lambda^{p-1}(M)$. Since variation and exterior
derivative commute with each other $ \delta d = d \delta $,
(\ref{85}) may be rewritten as
 \ba
   \delta I &=& \int_M { (d \delta A) \wedge (2* dA)
   - \delta e^a \wedge [(\imath_a F) \wedge * F - (-1)^p F \wedge (\imath_a
   *F)]} \nonumber \\
  &=& \int_M  (\delta A) \wedge (-1)^p (2 d*F) + d (\delta A \wedge 2 *
  F) \nonumber \\
  & & \quad \quad \quad \quad \quad - \delta e^a \wedge [(\imath_a F) \wedge * F - (-1)^p F \wedge (\imath_a
   *F)]
 \ea
By applying the Stoke's theorem, the second term on the right hand
side can be written as
 \ba
     \int_M d (\delta A \wedge 2 * F) = \int_{\partial M} \delta A \wedge 2 *
     F = 0
 \ea
because the boundary condition is $ \left. \delta A
\right|_{\partial M} =0$ where $\partial M$ is the boundary of
$M$. Thus
  \ba
   \delta I &=& \delta \int_M dA \wedge * dA \nonumber \\
    &=& \int_M  (\delta A) \wedge (-1)^p (2 d*F)
    - \delta e^a \wedge [(\imath_a F) \wedge * F - (-1)^p F \wedge (\imath_a
   *F)]
 \ea
where $F=dA \in \Lambda^p(M)$.

  \begin{center}
 \begin{tabular}{|c|}
   \hline
   {\it \large Special Case:2} \hskip 1cm $\alpha = e_a \wedge M^a$ and $\beta = e_a \wedge N^a$\\
   \hline
 \end{tabular}
 \end{center}

These kinds of Lagrangians are encountered in non-Riemannian
gravity models in which irreducible components of gauge fields are
used. In this case (\ref{deltaI2}) becomes
  \ba
   \delta I &=& \int_M  \delta (e_a \wedge M^a) \wedge * (e_b
   \wedge N^b) + \delta (e_a \wedge N^a) \wedge * (e_b \wedge M^b) \nonumber \\
   & & \quad \quad \quad - \delta e^a \wedge [\imath_a (e_b \wedge N^b) \wedge * (e_c \wedge M^c)
   -(-1)^p (e_c \wedge M^c) \wedge \imath_a * (e_b \wedge N^b)]
 \ea
where $M^a \; , \; N^a \in \Lambda^{p-1}(M)$. Organizing terms
 \ba
   \delta I &=& \int_M  \delta e_a \wedge M^a \wedge * (e_b \wedge N^b)
   + e_a \wedge \delta M^a \wedge * (e_b \wedge N^b) \nonumber \\
  & & + \delta e_a \wedge N^a \wedge * (e_b \wedge M^b)
   + e_a \wedge \delta N^a \wedge * (e_b \wedge M^b) \nonumber \\
   & & - \delta e^a \wedge [ N_a \wedge * (e_c \wedge M^c)
   - e_b \wedge \imath_a N^b \wedge * (e_c \wedge M^c) \nonumber \\
   & & -(-1)^p (e_c \wedge M^c) \wedge * (e_b \wedge N^b \wedge
   e_a)] \; .
 \ea
we obtain
  \ba
   \delta I &=& \delta \int_M  (e_a \wedge M^a) \wedge * (e_b \wedge N^b) \nonumber \\
   &=& \int_M \delta M^a \wedge e_a \wedge * (N^b \wedge e_b)
   + \delta N^a \wedge e_a \wedge * (M^b \wedge e_b) \nonumber \\
   & & \quad \quad \quad + \delta e^a \wedge [ M_a \wedge * (e_b \wedge N^b)
   + e_b \wedge (\imath_a N^b) \wedge * (e_c \wedge M^c) \nonumber \\
  & & \quad \quad \quad \quad \quad - (e_c \wedge M^c) \wedge * (e_a \wedge e_b \wedge N^b) ] \; .
 \ea

 \begin{center}
 \begin{tabular}{|c|}
   \hline
    {\it \large Special Case:3} \hskip 1cm $\alpha = \imath_a B^a$ and $\beta =\imath_a C^a$\\
   \hline
 \end{tabular}
 \end{center}

In this case (\ref{deltaI2}) becomes
  \ba
   \delta I &=& \int_M  \delta (\imath_a B^a) \wedge * \imath_b C^b
   + \delta (\imath_a C^a) \wedge * \imath_b B^b \nonumber \\
   & & \quad \quad \quad - \delta e^a \wedge [(\imath_a \imath_b C^b) \wedge * (\imath_c B^c)
   -(-1)^p (\imath_c B^c) \wedge (\imath_a *\imath_bC^b)]
   \label{deltaI3}
 \ea
where $B^a \; , \; C^a \in \Lambda^{p+1}(M)$. First we write the
first term on the right hand side as
 \ba
  \delta (\imath_a B^a) \wedge * \imath_b C^b = \imath_{\delta a} B^a \wedge * (\imath_b
  C^b)  + \imath_a \delta B^a \wedge * (\imath_b C^b) \; ,
  \label{deltaimathaBa}
 \ea
so that
 \ba
  \imath_{\delta a} B^a \wedge * (\imath_b C^b) &=& \imath_{\delta a}
    \left( \frac{1}{(p+1)!} {B^{a,}}_{i_1 \cdots i_{p+1}}
    e^{i_1 \cdots i_{p+1}}\right) \wedge * (\imath_b C^b) \nonumber \\
   &=& \frac{1}{(p+1)!} {B^{a,}}_{i_1 \cdots i_{p+1}}
   \left( \imath_{\delta a} e^{i_1 \cdots i_{p+1}} \right) \wedge * (\imath_b C^b) \nonumber \\
  &=& ( \imath_{\delta a} e^{i_1}) \wedge \frac{1}{p!} {B^{a,}}_{i_1 \cdots i_{p+1}}
    e^{i_2 \cdots i_{p+1}}  \wedge * (\imath_b C^b) \; .
 \ea
By making the use of the duality
 \ba
   \imath_a e^b = \delta ^b_a \quad \rightarrow \quad \delta (\imath_a
   e^b)=0 \quad \rightarrow \quad \imath_{\delta a} e^b = - \imath_a \delta e^b
 \ea
this equation takes the form
 \ba
  \imath_{\delta a} B^a \wedge * (\imath_b C^b)
 = -( \imath_a \delta e^{i_1}) \wedge \frac{1}{p!} {B^{a,}}_{i_1 \cdots i_{p+1}}
    e^{i_2 \cdots i_{p+1}}  \wedge * (\imath_b C^b) \; .
 \ea
Any $(n+1)$-form is equal to zero on a $n$-dimensional manifold,
i.e.,
  \ba
  \imath_a \left( - \delta e^{i_1} \wedge \frac{1}{p!} {B^{a,}}_{i_1 \cdots i_{p+1}}
    e^{i_2 \cdots i_{p+1}}  \wedge * (\imath_b C^b) \right) = 0 \; .
 \ea
This implpies
 \ba
   \imath_{\delta a} B^a \wedge * (\imath_b C^b)
 &=& (-1)^{p+1} \delta e^{i_1} \wedge \frac{1}{p!} {B^{a,}}_{i_1 \cdots i_{p+1}}
    e^{i_2 \cdots i_{p+1}}  \wedge  \imath_a * (\imath_b C^b) \nonumber \\
 & & \quad \quad \quad - \delta e^{i_1} \wedge \frac{1}{(p-1)!} {B^{a,}}_{i_1 a i_3 \cdots i_{p+1}}
    e^{i_3 \cdots i_{p+1}}  \wedge * (\imath_b C^b) \nonumber \\
 &=& (-1)^{p+1} \delta e^a \wedge (\imath_a B^c) \wedge  \imath_c * (\imath_b C^b)
    - \delta e^a \wedge (\imath_c \imath_a B^c) \wedge * (\imath_b C^b) \nonumber \\
 &=& \delta e^a \wedge [ (\imath_a \imath_c B^c) \wedge * (\imath_b C^b)
    - (\imath_a B^c) \wedge * ( e_c \wedge \imath_b C^b)]
    \label{4.1}
 \ea
where we used the identity $\imath_a *\Phi = *(\Phi \wedge e^a)$
for $\Phi \in \Lambda^p(M)$. Now we calculate the second term on
the right hand side of (\ref{deltaimathaBa}) by forming a
$(n+1)$-form on $n$-dimensional manifold
 \ba
     \imath_a [\delta B^a \wedge * (\imath_b C^b)] &=&0 \nonumber \\
   \imath_a \delta B^a \wedge * (\imath_b C^b) &=& \delta B^a \wedge * (e_a \wedge \imath_b
   C^b) \; . \label{4.2}
 \ea
Then collecting (\ref{4.1}) and (\ref{4.2}), we write down
(\ref{deltaimathaBa})
 \ba
  \delta (\imath_a B^a) \wedge * \imath_b C^b = \delta B^a \wedge * (e_a \wedge \imath_b C^b)
   + \delta e^a \wedge [ (\imath_a \imath_c B^c) \wedge * (\imath_b C^b)
    - (\imath_a B^c) \wedge * ( e_c \wedge \imath_b C^b)] \; . \label{deltaimath}
  \ea
In order to calculate the second term on the right hand side of
(\ref{deltaI3}) we just replace $B$ with $C$ in equation
(\ref{deltaimath}):
  \ba
  \delta (\imath_a C^a) \wedge * \imath_b B^b = \delta C^a \wedge * (e_a \wedge \imath_b B^b)
   + \delta e^a \wedge [ (\imath_a \imath_c C^c) \wedge * (\imath_b B^b)
    - (\imath_a C^c) \wedge * ( e_c \wedge \imath_b B^b)] \; . \label{deltaimath1}
  \ea
Thus, when we put all the results into (\ref{deltaI3}), we find
 \ba
 \delta I &=& \delta \int_M (\imath_a B^a) \wedge * (\imath_b C^b) \nonumber \\
   &=& \int_M \delta B^a \wedge * (e_a \wedge \imath_b C^b)
   + \delta C^a \wedge * (e_a \wedge \imath_b B^b) \nonumber \\
  & & \quad \quad  + \delta e^a \wedge [ (\imath_a \imath_c B^c) \wedge * (\imath_b C^b)
    - (\imath_a B^c) \wedge * ( e_c \wedge \imath_b C^b)
     - (\imath_a C^c) \wedge * (e_c \wedge \imath_b B^b) \nonumber \\
   & & \quad \quad \quad \quad \quad \quad \quad \quad
      \quad \quad + (\imath_c B^c) \wedge * ( e_a \wedge \imath_b C^b)]
  \ea
where $B^a \; , \; C^a \in \Lambda^{p+1}(M)$.

 \begin{center}
 \begin{tabular}{|c|}
   \hline
   {\it \large Special Case:4} \hskip 1cm $\alpha = \imath_a Q^{ab}$ and $\beta =\imath_b R$ \\
   \hline
 \end{tabular}
 \end{center}

In this case (\ref{deltaI2}) becomes
  \ba
   \delta I &=& \int_M  \delta (\imath_a Q^{ab}) \wedge * \imath_b R
   + \delta (\imath_b R) \wedge * \imath_b Q^{ab} \nonumber \\
   & & \quad \quad \quad - \delta e^a \wedge [(\imath_a \imath_b R) \wedge * (\imath_c B^{bc})
   -(-1)^p (\imath_c Q^{bc}) \wedge (\imath_a *\imath_b R)]
   \label{deltaI4}
 \ea
where $Q^{ab} \; , \; R \in \Lambda^{p+1}(M)$. First we handle the
first and the second terms on the right hand side
 \ba
  \delta (\imath_a Q^{ab}) \wedge * \imath_b R &=& \imath_{\delta a} Q^{ab}
  \wedge * (\imath_b R)  + \imath_a \delta Q^{ab} \wedge * (\imath_b R)\\
   &=& \delta Q^{ab} \wedge *(e_a \wedge \imath_b R) + \delta e^a
   \wedge [(\imath_a \imath_b Q^{bc}) \wedge *(\imath_c R) \nonumber \\
 & &  \quad \quad \quad \quad \quad - (\imath_a Q^{bc}) \wedge *(e_b \wedge \imath_c R)] \; , \\
  \delta (\imath_b R) \wedge * (\imath_a Q^{ab}) &=& \imath_{\delta a} R
  \wedge * (\imath_b Q^{ab})  + \imath_b \delta R \wedge * (\imath_a Q^{ab})\\
   &=& \delta R \wedge *(e_a \wedge \imath_b Q^{ab}) + \delta e^a
   \wedge [(\imath_a \imath_b R) \wedge *(\imath_c Q^{bc}) \nonumber \\
 & &  \quad \quad \quad \quad \quad  - (\imath_a R) \wedge *(e_b \wedge \imath_c Q^{bc})] \; .
 \ea
Thus
 \ba
 \delta I &=& \delta \int_M (\imath_a Q^{ab}) \wedge * (\imath_b R) \nonumber \\
   &=& \int_M \delta Q^{ab} \wedge * (e_a \wedge \imath_b R)
   + \delta R \wedge * (e_a \wedge \imath_b Q^{ab}) \nonumber \\
  & & \quad \quad  + \delta e^a \wedge [ (\imath_a \imath_b Q^{bc}) \wedge * (\imath_c R)
    - (\imath_a Q^{bc}) \wedge * ( e_b \wedge \imath_c R)
     - (\imath_a R) \wedge * (e_b \wedge \imath_c Q^{bc}) \nonumber \\
   & & \quad \quad \quad \quad \quad \quad \quad \quad
      \quad \quad + (\imath_c Q^{bc}) \wedge * ( e_a \wedge \imath_b R)]
  \ea
where $Q^{ab} \; , \; R \in \Lambda^{p+1}(M)$.


\begin{thebibliography}{99}
\bibitem{adak2003} M Adak, T Dereli and L H Ryder, {\it Int. J. Mod. Phys.} {\bf D12} (2003) 145-156 \\
                    (arXiv:gr-qc/0208042)
\bibitem{dereli1987} T Dereli and R W Tucker, {\it Class. Quant. Grav.} {\bf 4} (1987) 791
\bibitem{dereli1995} T Dereli and R W Tucker, {\it Class. Quant. Grav.} {\bf 12} (1995) L31
\bibitem{hehl1995}  F W Hehl, J D McCrea, E W Mielke and Y Ne'eman, {\it Phys. Rep.} {\bf 258} (1995) 1
\bibitem{tucker1995} R W Tucker and C Wang, {\it Class. Quant. Grav.} {\bf 12} (1995) 2587-2605
\bibitem{dereli1996} T Dereli, M \"{O}nder, J Schray, R W Tucker and C
                        Wang, {\it Class. Quant. Grav.} {\bf 13} (1996) L103-L110\\ (arXiv:gr-qc/9604039)

\bibitem{adak2004} M Adak, T Dereli and L H Ryder {\it Phys. Rev.} {\bf D69} (2004) 123002 \\ (arXiv:gr-qc/0303080)
\bibitem{dereli1982} T Dereli and R W Tucker, {\it J. Phys.} {\bf A15} (1982) 1625
\bibitem{adak2001} M Adak, T Dereli and L H Ryder {\it Class. Quant. Grav.} {\bf 18} (2001) 1503-1512 \\
                        (arXiv:gr-qc/0103046)
\bibitem{hayashi1967} K Hayashi and T Nakano, {\it Prog. Theor. Phys.} {\bf 38} (1967) 491
\bibitem{hayashi1979} K Hayashi and T Shirafuji, {\it Phys. Rev.} {\bf D19} (1979) 3524
\bibitem{obukhov2003} Y N Obukhov and J G Pereira, {\it Phys. Rev.} {\bf D67} (2003) 044016 \\ (arXiv:gr-qc/0212080)
\bibitem{maluf2003}  J W Maluf, {\it Phys. Rev.} {\bf D67} (2003) 108501 \\ (arXiv:gr-qc/0304005)
\bibitem{arcos2004}  H I Arcos, V C De Andrade and J G Pereira, {\it Int. J. Mod. Phys.} {\bf D13} (2004) 807 \\
                     (arXiv:gr-qc/0403074)
\bibitem{dereli1994} T Dereli and R W Tucker, {\it Class. Quant. Grav.} {\bf 11} (1994) 2575-2583
\bibitem{hehl2005} C Heinicke, P Baekler and F W Hehl, (arXiv:gr-qc/0504005)
\bibitem{nester1999}  J M Nester and H J Yo, {\it Chinese J. Phys.} {\bf 37} (1999) 113 \\ (arXiv:gr-qc/9809049)
\bibitem{adak2005}   M Adak and \"{O} Sert, {\it Turk. J. Phys.} {\bf 29} (2005) 1-7 \\ (arXiv:gr-qc/0412007)
\bibitem{vasilic2000} M Blagojevi\'{c} and M Vasili\'{c}, {\it Class. Quant. Grav.} {\bf 17} (2000) 3785 \\
                     (arXiv:hep-th/0006080)
\bibitem{obukhov1997} Y N Obukhov, E J Vlachynsky, W Esser and F W Hehl, {\it Phys. Rev.} {\bf D56} (1997) 7769
\bibitem{tucker1998}  R W Tucker and C Wang, {\it Class. Quant. Grav.} {\bf 15} (1998) 933-954
\bibitem{dereli2000} T Dereli, M \"{O}nder, J Schray, R W Tucker and C Wang {\it Non-Riemannian Gravitational
                        Interactions: An Overview} In the Proceedings of $9^{th}$ Marcel
                        Grossmann Meeting, July 2000, Rome, Italy



\end{thebibliography}
\end{document}